\documentstyle[11pt,newpasp,twoside,epsf]{article}
\def\edcomment#1{\iffalse\marginpar{\raggedright\sl#1\/}\else\relax\fi}
\reversemarginpar

\begin{document}
\title{Search for Nemesis Encounters with Vega, $\epsilon$ Eridani,
and Fomalhaut}
 \author{Jean-Marc Deltorn}
\affil{Space Telescope Science Institute, 3700 San Martin Drive,
Baltimore, MD 21218}
\author{Paul Kalas}
\affil{University of California, 601 Campbell Hall,
Berkeley, CA 94720}

\begin{abstract}
We calculate the space motions of 21,497 stars to
search for close stellar encounters with Vega, 
$\epsilon$ Eridani and Fomalhaut during the past 10$^6$ yr. 
We discover that $\epsilon$ Eridani experienced three $<$2 pc encounters
over the past $10^5$ yr.  Within the uncertainties, $\epsilon$ Eridani
is having a close encounter with
Kapteyn's star near the present epoch, with a 42.2\% probability
that the closest approach distance is $<$1 pc.
Vega and Fomalhaut experienced four and 
six $<$2 pc encounters, respectively, over the past 10$^6$ yr.
Each had one encounter with a $\sim$2\% probability 
that the closest approach
distance is less than 0.5 pc.
These encounters will not directly influence the debris disks observed
around Vega, $\epsilon$ Eridani and Fomalhaut, but they may pass through
hypothetical Oort clouds surrounding these stars.
We find that two other Vega-like stars, HD 17848 and HD 20010, 
experienced rare, $<$0.1 pc stellar encounters that are more 
likely to directly
perturb their circumstellar disks. 

\end{abstract}

\section{Introduction}

Objects orbiting a star may be redistributed or ejected
over their lifetime due to the close passage of another 
star with the system.
``Nemesis encounters'' may produce interstellar comets (Weissman 1996), 
free-floating planets (Laughlin \& Adams 1998; Smith \& Bonnell 2001), 
comet showers (Heisler et al. 1987) 
and periodic extinctions (Davis et al. 1984).
Though 
nemesis encounters are short-lived and rare, these events
may manifest as relatively long-lived phenomena.
Comet showers may increase the dust content close to the star
and produce an observable signature that decays on dust destruction
timescales.
Also, the direct dynamical influence of the perturber on an extant debris
disk may generate significant asymmetries in the disk morphology
(Larwood \& Kalas 2001).  

Here we use data from the Hipparcos catalog and the 
Barbier-Brossat \& Figon (2000) catalog of stellar radial 
velocities to determine the U, V, W space velocities of 
21,497 stars and trace their trajectories 10$^6$ yr into the past.  
We search for the closest stellar encounters experienced by 
three nearby stars -- Vega, $\epsilon$ Eridani, and 
Fomalhaut -- that have asymmetric debris disks 
(Holland et al. 1998; Greaves et al. 1998).

\section{Search for Nemesis Encounters}

The sample selection and method are identical to recent work
that tests for  nemesis encounters with $\beta$ Pic
(Kalas, Deltorn \& Larwood 2001).  In Table 1 we show those
encounters with Vega, $\epsilon$ Eri, and Fomalhaut that have
closest approach distance, D$_{ca}<2$ pc.  
We use a Monte-Carlo routine to propagate the standard
deviations for the observables (position, proper motion, parallax,
and radial velocity), resulting in a spread of final results for D$_{ca}$
and the closest approach time, t$_{ca}$.  
Columns 2 and 3 in Table 1  
give the maxima and 1-$\sigma$ values
for these distributions, and Fig. 1 shows the spread 
of values as isocontours.  
The contours trace the confidence level for finding
a star at a specific location in the (D$_{ca}$,t$_{ca}$) plane.
For each encounter, Table 1 gives the probability of
closest approach at the 95.4\% confidence level, P$_{95.4\%}$,
for D$_{ca} <$ 1 pc and D$_{ca} <$ 0.5 pc (columns 5 and 6, respectively).
These probabilities are computed by dividing the 
distribution below  D$_{ca} <$ 1 pc and D$_{ca} <$ 0.5 pc by
the total distribution in the (D$_{ca}$,t$_{ca}$) plane (Fig. 1).

\begin{table}
\centering
\caption{Closest Approach Parameters for the Stars with $\bar{D}_{ca}<2$ pc}
\begin{tabular}[h]{lccccc}
\hline \hline\\
Hipparcos No. & t$_{ca}$              & D$_{ca}$   & $\Delta$V$_{ca}$   & P$_{95.4\%}^{(D_{ca}<1pc)}$  & P$_{95.4\%}^{(D_{ca}<0.5pc)}$\\
      	     & (kyr)                 & (pc)       & (km/s)             &       ($\%$)          &       ($\%$)      \\\hline
{\bf Vega:}	&			&		&		&		&	\\		
HIP 21421       & -620.1$^{+26.6}_{-26.4}$      &1.84$^{+0.68}_{-0.48}$ &38.6$\pm0.8$           & 0.0   &0.0    \\
HIP 34603       & -337.9$^{+29.5}_{-28.9}$      &1.50$^{+0.68}_{-0.61}$ &31.6$\pm2.1 $          & 17.8  &1.9    \\
HIP 85665       & -318.0$^{+36.5}_{-33.2}$      &1.49$^{+0.36}_{-0.33}$ &18.0$\pm1.4 $          & 1.0   &0.0    \\
HIP 86974       & -52.3$^{+10.1}_{-9.9}$        &1.08$^{+0.21}_{-0.20}$ &41.1$\pm0.3 $          & 31.0  &0.0    \\
\hline
{\bf $\epsilon$ Eri:}&			&		&		&		&			\\
HIP 24186       & -12.5$^{+12.2}_{-9.8}$        &1.02$^{+0.22}_{-0.20}$ &297.8$\pm1.0 $         & 42.2  & 0.0   \\
HIP 30920       & -62.3$^{+6.7}_{-6.6}$         &1.15$^{+0.18}_{-0.17}$ &41.9$\pm1.5 $          & 11.3  & 0.0   \\
HIP 32349       & -72.3$^{+3.3}_{-6.1}$         &1.83$^{+0.06}_{-0.13}$ &21.4$\pm0.2 $          & 0.0   & 0.0   \\
\hline
{\bf Fomalhaut:} &				&		&		&		&			\\
HIP 12777       & -474.0$^{+20.0}_{-19.1}$      &1.15$^{+0.41}_{-0.34}$ &28.8$\pm0.5 $          &21.8   & 0.0   \\
HIP 31626       & -490.6$^{+43.2}_{-46.5}$      &1.77$^{+0.61}_{-0.68}$ &76.8$\pm1.1 $          &10.5   & 1.6   \\
HIP 53767       & -394.3$^{+73.0}_{-89.6}$      &1.84$^{+0.54}_{-0.61}$ &34.5$\pm4.9 $          &1.4    & 0.0   \\
HIP 102485      & -281.4$^{+16.6}_{-13.2}$      &1.63$^{+0.34}_{-0.27}$ &30.3$\pm0.6 $          &0.0    & 0.0   \\
HIP 106440      & -185.2$^{+17.3}_{-15.8}$      &1.78$^{+0.47}_{-0.55}$ &17.9$\pm0.4 $          &1.2    & 0.0   \\
HIP 116745      & -102.1$^{+7.8}_{-7.5}$        &1.90$^{+0.18}_{-0.18}$ &75.3$\pm1.2 $          &0.0    & 0.0   \\
\hline
\end{tabular}
\end{table}

The dynamical influence of each encounter depends 
on the relative velocity ($\Delta$V$_{ca}$, Table 1) and mass (Table 2) of
each perturber.  Following Kalas et al. (2001), we use the impulse approximation
to compute the average change of
velocity, $\Delta v_{avg}$, and eccentricity, $\Delta e_{avg}$,
for comets in a hypothetical, 10$^5$ AU radius Oort cloud around
each star (Table 2).

\section{Discussion}

Our search for nemesis encounters shows that
Vega, $\epsilon$ Eri, and Fomalhaut experienced several $<$2 pc
encounters over the past 10$^6$ yr, 
comparable to the rate of encounters for
the Sun (Garc\'{\i}a-S\'anchez et al. 1999).  
The most recent encounters are interesting because
their dynamical influence on circumstellar disks may
still be evident in observations (e.g. Kalas et al. 2000).
Within the uncertainties,
$\epsilon$ Eri's D$_{ca}\sim$ 1 pc encounter with HIP 24186
(Kapteyn's star) is occurring at the present
epoch.  There is a 42.2\% probability that the Kapteyn-$\epsilon$ Eri encounter
has D$_{ca}<$ 1 pc (Table 1).  
However, the large relative velocity of this encounter
and the low mass of Kapteyn's star combine to give the smallest
dynamical influence in our study (Table 2).  

\begin{table}
\centering
\caption{Dynamical Influence on
Objects Orbiting at $10^5$ AU Radius}
\begin{tabular}[h]{lccccccc}
\hline \hline\\
Hipparcos No.&  Mass          & $\Delta v_{avg}$ &  $\Delta e_{avg}$ & R$_{eq}$\\
             &  (M$_{\odot}$) &(10$^{-3}$ m/s)  &   (10$^{-3}$)     & (pc)    \\
\hline 
{\bf Vega:}  &                  &               \\
21421    & 0.67  & 20.39$\pm 6.35$  &   0.25$\pm 0.08$ & 1.13$^{+0.42}_{-0.29}$ \\
34603    & 0.21  & 17.61$\pm10.31$  &   0.22$\pm 0.20$ & 1.11$^{+0.50}_{-0.45}$  \\  
85665    & 0.45  & 40.62$\pm 7.09$  &   0.51$\pm 0.09$ & 0.98$^{+0.24}_{-0.22}$ \\ 
86974   & 0.95 & 81.30$\pm 8.60$  &   1.01$\pm 0.11$  & 0.62$^{+0.12}_{-0.11}$\\ 
\hline 
{\bf $\epsilon$ Eri:}&
                    &  \\ 
24186    & 0.5  &  6.05$\pm 0.22$  &   0.15$\pm 0.01$ &0.57$^{+0.12}_{-0.11}$ \\ 
30920    & 0.2  & 14.17$\pm 1.78$  &   0.35$\pm 0.04$ &0.77$^{+0.12}_{-0.11}$ \\    
32349   & 2.9  &167.87$\pm 5.52$  &   4.20$\pm 0.14$  &0.63$^{+0.02}_{-0.04}$\\   
\hline 
{\bf Fomalhaut:}&                                       &\\
12777    & 1.3 &129.92$\pm22.14$  &   1.82$\pm 0.31$ &0.51$^{+0.10}_{-0.09}$\\ 
31626     & 0.67 & 17.16$\pm14.64$  &   0.24$\pm 0.20$ &0.92$^{+0.21}_{-0.18}$ \\ 
53767     & 0.3 & 10.22$\pm 0.79$  &   0.14$\pm 0.01$& 1.14$^{+0.38}_{-0.42}$\\
102485    & 1.4 & 68.60$\pm 9.25$  &   0.96$\pm 0.13$  &0.70$^{+0.15}_{-0.12}$ \\
106440    & 0.45 & 33.12$\pm16.46$  &   0.46$\pm 0.23$ &1.02$^{+0.27}_{-0.31}$\\ 
116745    & 0.7 & 10.57$\pm 1.05$  &   0.15$\pm 0.01$  &0.98$^{+0.09}_{-0.09}$\\    
\hline
\end{tabular}
\end{table}

The $\epsilon$ Eri encounters are distinguished by the
fact that all three occurred recently, t$_{ca}<$ 10$^5$ yr
(Vega and Fomalhaut each have one encounter with t$_{ca}\la$ 10$^5$ yr).
Also, $\epsilon$ Eri's encounter with HIP 32349 (Sirius) gives the strongest
dynamical influence on hypothetical Oort clouds in our group (Table 2).
These results partly reflect the fact that at the present epoch $\epsilon$ Eri
has a smaller heliocentric distance (3.3 pc) than Vega (7.8 pc) or 
Fomalhaut (7.7 pc).  As a star enters a volume with greater
observational completeness, the likelihood of finding recent close
encounters increases (Garc\'{\i}a-S\'anchez et al. 1999). 
However, Vega and Fomalhaut may have experienced the closest stellar encounters in
this study,
with a $\sim$2\% probability 
that D$_{ca} <$ 0.5 pc.
(Table 1, Fig. 1). 

None of these encounters will directly influence the observed 
circumstellar disks, which appear confined to $<$0.001 pc 
radius (Holland et al. 1998; Greaves et al. 1998).  
However, an undetected assembly of comets weakly bound to each star 
with semi-major axis $\sim$1 pc may experience a 
dynamical perturbation.  

Table 2 gives the distance, $R_{eq}$, where the gravitational forces 
between Vega, $\epsilon$ Eri and Fomalhaut, and each of their respective stellar 
perturbers, are equal at t$_{ca}$.  
The estimated values of $R_{eq}$ are less than the maximum 
radius of a gravitationally bound cloud of comets around each star. 
Stars that pass through an Oort cloud can erode the comet population
by ejecting members into interstellar space or by producing comet showers near the star 
that may briefly increase the circumstellar dust content (Weissman 1996).
Another source of perturbation for Fomalhaut is its probable binary stellar 
companion (Barrado y Navascues et al. 1997). 
In principle the close encounters identified
here may pump the orbital eccentricity of the companion such that Fomalhaut's
debris disk experiences periodic close encounters with the companion.

Establishing whether or not the dust debris around Vega, $\epsilon$ Eri,
and Fomalhaut is enhanced because of these recent $\sim$1 pc encounters requires
a comprehensive study using a control sample.  Two more Vega-like stars that
should also be studied closely are HD 17848 and HD 20010 (Sylvester \& Mannings 
2000; Walker \& Wolstencroft 1988).
We have discovered that HD 17848 and HD 20010 encountered each other with
D$_{ca}$ = 0.081 pc, t$_{ca}$ = -351.2 kyr, and $\Delta$V$_{ca}$ = 44.2 km s$^{-1}$.
The 1-$\sigma$ uncertainty for D$_{ca}$ is (-0.0488,+0.6250) pc.

Presently the greatest limitation in conducting these studies
lies with the radial velocity catalogs, which 
have less than a third of the stars in the Hipparcos catalog.
A definitive study of nemesis encounters will be possible after
an all-sky radial velocity survey, such as the one proposed for GAIA, is 
completed, and after
more sensitive astrometric catalogs become available, such as the one expected
in $\sim$2006 from the FAME mission.

\begin{figure}
\plotone{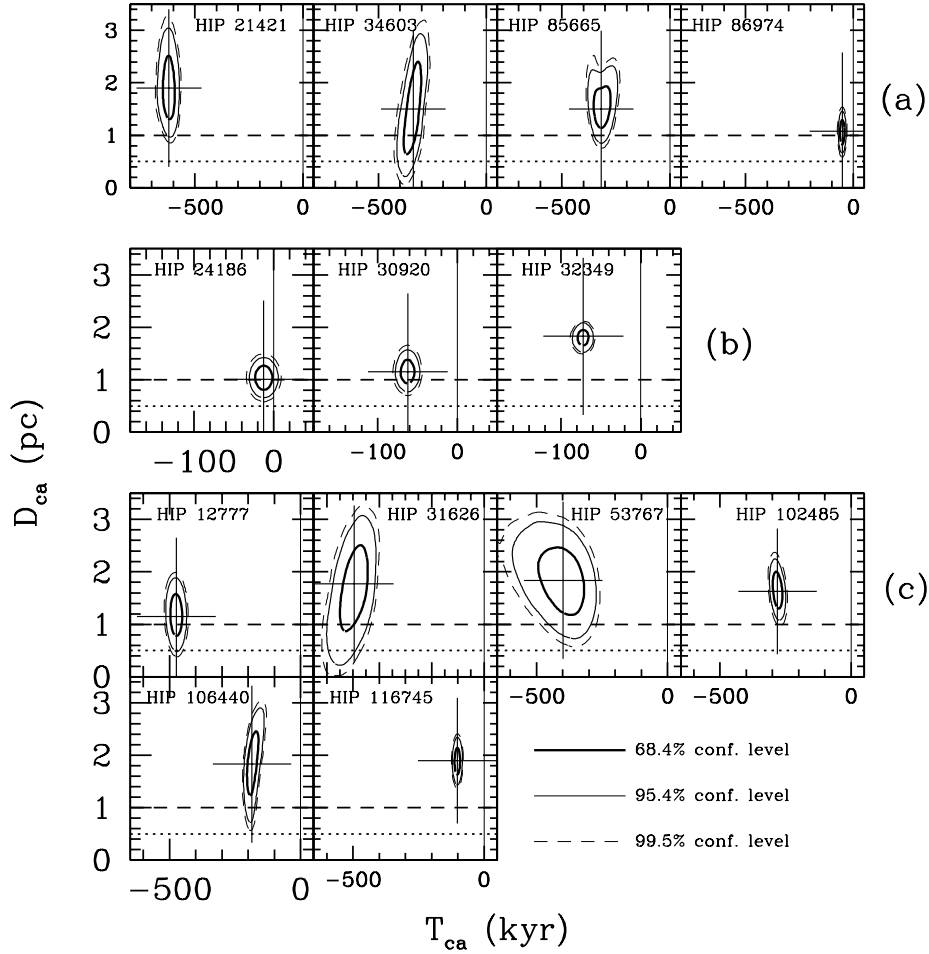}
\caption{Isocontours for the Monte-Carlo distributions
in the  closest approach plane,  (${D}_{ca},t_{ca}$),
for the stars with $\bar{D}_{ca}<2$ pc during the past 10$^6$ yr
relative to:  a) Vega, b) $\epsilon$ 
Eridani, and c) Fomalhaut.
Bold contour: 68.3$\%$ confidence level, thin contour: 
95.4$\%$, and dashed contour: 99.5$\%$  confidence level. 
The contours outline the decreasing
probability of finding the star at a more specific location at a given
time due to uncertainties in position, proper motion, radial velocity, and parallax.  
Each cross marks the maximum of the probability
distribution derived from the Monte 
Carlo simulation (the size of the cross has no significance).
For a fixed uncertainty in the observables, the area of 
contours will grow the farther back in time an encounter occurs.
Horizontal lines mark ${D}_{ca}$ = 1 pc and  ${D}_{ca}$ = 0.5 pc.
Note that the nemesis search extended to t$_{ca}$ = -1000 kyr, but
here we plot each x-axis only as far back in time as
required to encompass the oldest $<$2 pc encounter within
the 10$^6$ yr timeframe.
}
\end{figure}

\end{document}